\documentclass[11pt,a4paper]{JHEP}
\usepackage{epsfig,bbm}
\usepackage{amssymb} 

\def\ln{{\rm ln}}

\def\bX{\overline X}

\def\G{{\cal G}_{w\bar w}}
\def\Z{{\mathbbm Z}}
\def\d{\partial}
\def\dw{\partial_w}

\def\dbw{\partial_{\bar w}}

\def\sinh{{\rm sinh}}
\def\det{{\rm Det}}
\def\Tr{{\rm Tr}}
\newcommand\0{\nonumber}
\newcommand\ee{\end{eqnarray}}	 	
\newcommand\be{\begin{eqnarray}}
\newcommand\ba{\begin{array}}			
\newcommand\ea{\end{array}}

\newcommand\e{{\rm e}}

\preprint{SISSA/85/98/EP\\\tt hep-th/9807232}

\title{String Interactions from Matrix String Theory}

\author{G.\ Bonelli, L.\ Bonora, F.\ Nesti\\
International School for Advanced Studies (SISSA/ISAS)\\
Via Beirut 2--4, 34014 Trieste, Italy, and INFN, Sezione di Trieste\\
E-mail: \email{bonelli@sissa.it}, \email{bonora@frodo.sissa.it}, 
\email{nesti@frodo.sissa.it}}

\abstract{The Matrix String Theory, i.e. the two dimensional U(N) SYM with
${\cal N}=(8,8)$ supersymmetry, has classical
BPS solutions that interpolate between an initial and a final string 
configuration via a bordered Riemann surface. The Matrix String Theory
amplitudes around such a classical BPS background, in the strong Yang--Mills 
coupling, are therefore candidates to be interpreted in a stringy way as 
the transition amplitude between given initial and final string 
configurations. In this paper we calculate
these amplitudes and show that the leading contribution is proportional
to the factor $g_s^{-\chi}$, where $\chi$ is the Euler characteristic of the
interpolating Riemann surface and $g_s$ is the string coupling. This is
the factor one expects from perturbative string interaction theory.
\\\\
{\sc Pacs:} 11.15.-q, 11.25.Sq, 11.27.+d}


\begin{document}

\section{Introduction}

The ${\cal N}=(8,8)$ SYM on a cylindrical 2D space--time with gauge group U(N) 
(hereafter referred to as Matrix String Theory (MST)) is expected to 
represent in the strong coupling limit a theory of type II
superstrings~\cite{motl,BS,WT}. 
In the naive strong coupling limit the action reduces to the 
Green--Schwarz action for free closed strings of various lengths. 
More than that, it has been suggested that this theory describes a second 
quantized superstring theory~\cite{DVV} (see also~\cite{DMVV,HV,BC} and the 
review article~\cite{DVVr}). 
In a couple of recent papers,~\cite{GHV,bbn}, it has been pointed out 
that the MST contains BPS instanton solutions which interpolate
between different
initial and final string configurations. In our previous paper,~\cite{bbn}, 
we remarked that this could be the clue for a comparison with
perturbative string interaction theory, not only from a qualitative, but also
from a quantitative point of view. This is what we intend to develop in this
paper.

The main aim of this paper is to show that the MST in the strong coupling limit
in the background of a given classical BPS instanton solution reduces to the
Green--Schwarz superstring theory plus a decoupled Maxwell theory, and
to compute amplitudes in such background. Since the latter interpolates
between an initial and a final string configuration via a 
bordered Riemann surface $\Sigma$ (which represents a branched covering of the
base cylinder), the amplitudes can be 
interpreted in a stringy way as the transition amplitudes between  
two such configurations. We show that their leading term 
is proportional to $g_s^{-\chi}$, where $\chi=2-2h-b$ is the Euler
characteristic of the Riemann surface of genus $h$ with $b$
boundaries, which characterizes the given classical
solution. This is the result one expects from perturbative string 
interaction theory.

The above derivation is contained in section~\ref{sec:2} and is organized as follows.
We first review the salient features of MST. Then we set out to compute the
strong YM coupling limit $g$. First we compute it for a classical BPS
background (part of the elaboration is contained in Appendix~\ref{A}). Then
we expand the action about this classical configuration by splitting
the quantum modes in two sets: the Cartan and the non--Cartan modes.
It turns out that the non--Cartan modes can be neatly integrated out 
(Appendix~\ref{B}). What remains is a quadratic action for the diagonal (Cartan) 
modes. However these modes are not individually well defined fields on the 
cylinder. A field 
interpretation is possible if we lift them to the covering $\Sigma$. We show 
that, if we do so, we obtain the Green--Schwarz theory plus the free Maxwell 
theory on the world--sheet $\Sigma$. Afterwards, we pass to the calculation
of the partition function of this theory and of the amplitudes 
mentioned above. In this regards a fundamental role is played by the Maxwell
field zero modes. With a careful computation one can show that this
amplitude is proportional to the factor $g^\chi= g_s^{-\chi}$ announced above.

\section{The Matrix String Theory and its strong coupling limit}
\label{sec:2}

\subsection{Euclidean MST and the instanton background}\label{sec:21}

To start with let us summarize the results of \cite{bbn}. MST is a theory 
defined
on a cylinder ${\cal C}$ with coordinates $\sigma$ and $\tau$. Its Euclidean 
action is
\begin{eqnarray} 
S&=&\frac{1}{\pi} \int_{\cal C} d^2w \,\Tr \Bigg(
D_w X^i D_{\bar w} X^i - \frac{1}{4g^2} F_{w\bar w}^2 -
\frac{g^2}{2}[X^i,X^j]^2 \0\\
&&~~~~~~~~~~~~~~~~~~~ +i (\theta^-_s D_{\bar w} \theta^-_s + \theta^+_c
D_w \theta^+_c) + ig \theta^T \Gamma_i [X^i,\theta] \Bigg),\label{eSYM}
\end{eqnarray}
 where we use the notation 
\begin{eqnarray}
w= \frac {1}{2} (\tau +i \sigma),\quad \bar w = \frac {1}{2} (\tau - i \sigma),
\quad\quad A_w= A_\tau-iA_\sigma ,\quad A_{\bar w}= A_\tau+iA_\sigma\,.\0
\end{eqnarray}
 
 Moreover $X^i$ with $i=1,\ldots,8$ 
are hermitean $N\times N$ matrices and
$D_w X^i = \partial_w X^i + i[A_w, X^i]$. $F_{w\bar w}$ is the gauge curvature.
Summation over the $i,j$ indices is understood. 
$\theta$ represents 16 $N\times N$ matrices whose entries are 2D 
spinors. It can be written as $\theta^T= (\theta^-_s,\theta^+_c)$, 
where $\pm$ denotes the 2D chirality and $\theta^-_s,\theta^+_c$ 
are spinors in the ${\bf 8_s}$ and ${\bf 8_c}$ representations of $SO(8)$, 
while $^T$ represents the 2D transposition.  The matrices $\Gamma_i$ are the 
$16\times 16$ $SO(8)$ gamma matrices. For definiteness we will write them as
\begin{equation}
\Gamma_i = \left(\matrix {0 & \gamma_i\cr
                 \tilde \gamma_i & 0\cr} \right),\0
\end{equation}
and $\gamma_i,\tilde\gamma_i$ are the same as in Appendix 5B of~\cite{GSW}.

The action (\ref{eSYM}) has ${\cal N}=(8,8)$ supersymmetry.  
In \cite{bbn} we singled out classical supersymmetric configurations
that preserve a $(4,4)$ supersymmetry. 
In this configurations the fermions are zero, $\theta=0$, and $X^i =0$ for all 
$i$ except two, for definiteness $X^i\neq 0$ for 
$i=1,2$. Introducing the complex notation $X=X^1+iX^2$, $~\bar X=
X^1-iX^2= X^\dagger$, the conditions to be satisfied for such BPS
configurations are
\footnote{Notice that one can obtain
anti-instantonic configurations by choosing an opposite polarization 
for the $(4,4)$ broken supersymmetries.} 
\begin{eqnarray}
&&F_{w\bar w} + i g^2 [X, \bar X] =0\label{insteq1}\\
&&D_w X=0, \quad\quad D_{\bar w} \bar X=0\,.\label{insteq2}
\end{eqnarray}

In \cite{bbn} we found explicit solutions of these equations.
From a mathematical point of view, (\ref{insteq1}, \ref{insteq2}) can be
identified with a {\it Hitchin system}~\cite{hitchin1} on a cylinder.
Each solution of (\ref{insteq1})(\ref{insteq2}) consists of two parts: a 
branched 
covering of the cylinder via the relative $X$ characteristic polynomial and 
a `dressing' 
factor. 

This can be seen by parametrizing the solution as
$X = Y^{-1}M Y $ and $A_w = -i Y^{-1}\d_w Y$, where $Y$ takes values in 
the complex
group $SL(N,{\mathbb C})$ \footnote{See Appendix~\ref{A} for the change from  
$GL(N,{\mathbb C})$ to  $SL(N,{\mathbb C})$ with respect to \cite{bbn}} 
and the matrix $M$ will be explained below. 
The dressing factor is contained in $Y$ while the branched covering is 
determined by $M$. 

Now, our purpose is to expand the action (\ref{eSYM}) around a generic 
classical BPS solution in inverse powers of the YM coupling $g$. Since, 
clearly, the background depends on $g$, we have to discuss preliminarily the 
strong coupling limit of the background itself.

Let us review the branched covering part first. We consider the polynomial
$$P_X(\mu)=\det (\mu - X)=\mu^N+\sum_{i=0}^{N-1}\mu^ia_i,$$
where $\mu$ is a complex indeterminate.
Due to (\ref{insteq2}), we have $\dw a_i=0$ which means that 
the set of functions $\{a_i\}$ are antianalytic on the cylinder.
Therefore the equation $$P_X(\mu)=0$$ identifies in the
$(w,\mu)$ space a Riemann surface $\Sigma$, which is an N--sheeted branched 
covering of the cylinder. The explicit form of the covering map is given 
by the X eigenvalues set $\{x^{(1)}(w),\dots,x^{(N)}(w)\}$.
The branch points are the locus where two or more eigenvalues coincide,
which means where the identification cuts in the sheets start or end. 
As for the 
parametrization of the branched coverings we will choose the standard one
\be
M=\left(\matrix{-a_{N-1}& -a_{N-2}& \ldots & \ldots  & -a_0\cr
		  1      & 0       & \ldots & \ldots  & 0  \cr
		  0       & 1       &  0  & \ldots  & 0 \cr
		  \ldots	   & \ldots     &  \ldots& \ldots  & 
		  \ldots \cr		
                  0	    & 0       & \ldots & 1    & 0 \cr
}\right).\label{M}
\ee
Notice that the branched covering structure is completely encoded in the 
$\{a_i\}$ analytic functions and {\it is independent of the value of the 
coupling}. 

The dependence on the coupling is entirely contained in the 
factor $Y$. Unfortunately we do not have yet an explicit analytical 
treatment of the way the strong coupling configuration is reached by
this factor, valid for any kind of covering. Therefore we limit ourselves to 
outline its features
in general, extrapolating the validity of our discussion from the example of 
the $\Z_N$ coverings, which we are able to deal with more effectively, 
\cite{bbn}.
In Appendix~\ref{A} we work out some examples in detail. From those results
it is sensible to assume that $Y$ tends,  in the strong coupling limit, to a 
precise matrix $Y_s$ (independent of $g$). This conclusion needs a more 
precise statement: if we cut out a neighborhood of size proportional to 
some positive power of $1/g$ 
around each branch point, then $YY_s^{-1}= Y_d$ dies off to 1, outside such a 
neighborhood, more quickly than any inverse powers of $g$. $Y_d$ will be 
called the dressing factor. Since in this paper we are interested in expanding 
the action (\ref{eSYM}) in inverse powers of $1/g$, and actually in singling 
out the dominant term in this expansion (see below), we will consider the 
action (\ref{eSYM}) around a given classical solution stripped of the above 
dressing factor and exclude from the integration region the branch points on 
the cylinder. In other words we will consider from now on the action 
(\ref{eSYM}) in which the pertinent $Y$ is replaced by 
$Y_s$ and the integral extends over ${\cal C}_0$ which is the initial cylinder 
${\cal C}$ from which the branch points have been removed. In the 
${\mathbb Z}_N$ case the surviving factor $Y_s$ is $( \sqrt{J \bar J})^{-1}$, 
with the notation of  \cite{bbn}.

After getting rid of the dressing factor, the classical background 
configuration is specified by 
$X = Y_s^{-1}M Y_s $ and $A_w = -i Y_s^{-1}\d_w Y_s$. As expected, this 
configuration is singular exactly at the branch points. Now one can 
diagonalize the matrix $M$ (\ref{M}), $M = S\hat X S^{-1}$, with a nonsingular 
matrix 
$S$ which can always be decomposed as
the product of a Hermitean matrix $W$ and a unitary matrix $T$, 
$S= WT$,
\cite{wynter},  (in the ${\mathbb Z}_N$ covering case, $W= J^{-1}$ and 
$T= \Lambda^{-1}$, see \cite{bbn}). $\hat X$ is the matrix of eigenvalues of 
$X$ and $X = Y_s^{-1} S \hat X S^{-1} Y_s$ in the strong coupling limit. 
Remarkably, $ Y_s^{-1} S $ is itself a (singular) unitary matrix and so 
simultaneously diagonalizes $X$ and $\bar X$. Corresponding to $\hat X$ 
we have $\hat A_w= -i S^{-1}\d_w S$ which can be seen to be zero everywhere, 
even at the branch points. What has happened is that the unitary 
transformation has swallowed entirely the connection, including the 
singularities.

We have therefore two distinguished (but equivalent) ways to represent the 
classical background 
in the strong coupling limit:
\begin{itemize}
\item strong coupling representation: $X_s= Y_s^{-1}M Y_s $ and $A_s 
= -i Y_s^{-1}\d Y_s$,
\item diagonal representation: $\hat X$ is diagonal and $\hat A=0$.
\end{itemize} 
The first is the natural form of the background on the cylinder once the 
dressing factor is removed by the strong coupling limit. The second, as we 
will see later on, is suitable to represent the theory on the covering 
space $\Sigma$.

To end this subsection we make two comments. The first concerns
the consistency of our procedure. One may feel uneasy for the appearance of
singularities and the use of singular gauge transformations, as above. 
As stressed in 
\cite{bbn}, the classical configuration specified by a given couple $(X, A)$ 
is smooth on the initial cylinder ${\cal C}$, but if we strip the solution 
of the dressing factor we get a configuration which is singular exactly at 
the branch points. The dressing factor is there exactly to compensate for 
these singularities. As pointed out in Appendix~\ref{A}, $Y_d-1$ has support only 
at the branch points in the strong coupling limit. It is perhaps possible, but 
formally very complicated, to keep such factor in the action. In this paper 
we prefer to replace ${\cal C}$ with ${\cal C}_0$ by excluding the branch 
points from the integration region to preserve smoothness. This is justified by
the following consideration. Beside the initial smooth configuration $(X, A)$
on ${\cal C}$, we will meet another smooth situation when we lift our theory
to the branched covering $\Sigma$ of ${\cal C}$ (see below). At that stage
the branch points can safely be restored in the integration region. What has
happened is that, in order to pass to the covering, we need a singular gauge 
transformation, the $ Y_s^{-1} S $ used above, which exactly kills the
singularity exposed by the strong coupling limit. In other words, it is 
natural to perform this singular gauge transformation if we want to reach a 
smooth situation which is fit for field theory.

We can regard the same problem from a different perspective. The
moduli space of couples $(X,A)$ satisfying (\ref{insteq1},\ref{insteq2}) have
been studied with sophisticated mathematical methods \cite{hitchin1}. However
we can say, roughly speaking, that they consist of the group degrees of freedom
(the factor $Y$ and their generalizations, modulo gauge transformations) times
the moduli space of the Riemann surfaces determined by the branched coverings.
The strong coupling limit suppresses the former and only the latter survive.
Since the risk of using singular gauge transformations is to suppress degrees of
freedom or to introduce new ones, we see that in our case this does not happen:
the moduli present in the theory after lifting it to the covering will 
correspond to those that have not been suppressed by the strong coupling
limit. 

The second comment concerns another way to get the strong coupling limit of 
the background; here we simply outline it.
In the parametrization
$$A_w=-iY^{-1}\dw Y\,\, ,\quad X=Y^{-1}MY\,,
$$
we get
$$
F_{w\bar w}\equiv \dw A_{\bar w}-\dbw A_w +i\left[A_w,A_{\bar w}\right]=
iY^{-1}\dw\left[\dbw\left(YY^+\right)\left(YY^+\right)^{-1}\right]Y
$$
and
$$
\left[X,\bar X\right]=Y^{-1}\left[M,\left(YY^+\right)M^\dagger
\left(YY^+\right)^{-1}\right]Y\,\, .$$
Therefore the BPS equation on $Y$ is
$$
\dw\left(\dbw\Omega\Omega^{-1}\right)+g^2\left[M,\Omega 
M^\dagger\Omega^{-1} \right]=0\, ,$$
where $\Omega\equiv YY^+$.
Notice that this equation is the extremality condition for the deformed 
WZNW action
$$ \qquad 
I_g\left[\Omega\right]=
\frac{1}{16\pi}\int d^2w \Tr \Omega^{-1}\dw\Omega\Omega^{-1}\dbw\Omega+
$$ $$
+\frac{1}{24\pi}\int_{B|\partial B=Cyl.}d^3y \varepsilon^{ijk}\Tr
\left(\Omega^{-1}\partial_i\Omega\Omega^{-1}\partial_j\Omega
\Omega^{-1}\partial_k\Omega\right)
-\frac{g^2}{4\pi}\int d^2w\Tr\left(\Omega^{-1}M\Omega M^+\right)\, .
\qquad$$
One could follow the $g$ deformation of the space of
classical solutions of the above theory subject to appropriate boundary
conditions and obtain its large $g$ limit analytically.

\subsection{Fixing the gauge and integrating along the non--Cartan 
directions}\label{2.2}

To extract the strong YM coupling effective theory, we first rewrite 
the action in the following useful form
\be
S=\frac{1}{\pi} \int d^2w \,&\Tr& \left(
D_w X^I D_{\bar w} X^I 
-\frac{g^2}{2}[X^I,X^J]^2
-g^2[X^I,X][X^I,\bX]
+D_w X D_{\bar w} \bX 
\right.\0\\
&& \left.
-\frac{1}{4g^2}\left( F_{w\bar w} +i g^2[X,\bX]\right)^2
 +i (\theta^-_s D_{\bar w} \theta^-_s + \theta^+_c
D_w \theta^+_c) + ig \theta^T \Gamma_i [X^i,\theta] \right)\0\,,
\ee
where $I= 3,4,...,8$. 
We now expand the action around a generic instanton configuration 
writing any field $\Phi$ as
\be
\Phi=\Phi^{(b)}+\phi^{\mathfrak t}+\phi^{\mathfrak n}\equiv 
\Phi^{(b)}+\phi\equiv \Phi^\circ +\phi^{\mathfrak n}\,,
\ee
where $\Phi^{(b)}$ is the background value of the field at infinite
coupling, $\phi^{\mathfrak t}$ are the fluctuations along the Cartan directions
and $\phi^{\mathfrak n}$ are the fluctuations 
along the complementary directions in Lie algebra ${\mathfrak u}(N)$. Of course
only the upper case fields $X,\bar X$ and $A$ will have non 
zero background value. By their background value we mean either the 
strong coupling or the diagonal representation introduced above. In the latter 
case $\phi^{\mathfrak t}$ and $\phi^{\mathfrak n}$ are the diagonal and 
non-diagonal part of $\phi$, respectively. In the strong coupling 
representation $\phi^{\mathfrak t}$ is the part of $\phi$ such that 
$\Tr (\phi X^n_s)\neq 0$ for some $n$, while $\phi^{\mathfrak n}$ is the part 
of $\phi$ for which $\Tr (\phi X^n_s)= 0$ for any power $n>0$. Anything we say 
in this subsection holds for both representations.

The first important remark is that, since the value of the action on the 
background is zero and since the background is a solution of the equation 
of motion, the 
expansion of the action starts with quadratic terms in the fluctuations.
To proceed further we have to fix the gauge.
We use, in the strong coupling limit, a gauge fixing inspired by
t'Hooft `abelian' gauges, and similar to the one used 
in \cite{chico}. We write it in the form
\be
\G=D^\circ_w a_{\bar w} +D^\circ_{\bar w} a_w + i g^2 ([X^\circ, \bar x]
+ [{\bar X}^\circ, x])+ 2i g^2 [ X^{\circ I}, x^I]  =0\label{gf}\,,
\ee
where $D^\circ$ stands for the covariant derivative with respect to $A^\circ$.
Next we introduce the Faddeev--Popov ghost and antighost fields $c$ and 
$\bar c$ and expand them like all the other fields.
Then we add to the action the gauge fixing term
\be
S_{gf}=\frac{1}{4\pi g^2}\int d^2w~\G^2
\ee
and the corresponding Faddeev--Popov ghost term
\be
S_{ghost}= -\frac{1}{2\pi g^2} \int d^2w ~\bar c \frac {\delta \G}{\delta c} c\,,
\ee
where $\delta$ represents the gauge transformation with parameter $c$.

At this point, to single out the strong coupling limit of the action,
we rescale the fields in appropriate manner. Precisely, we redefine our fields 
as follows
\be
A_w = A^{(b)}_w + g a_w^{\mathfrak t} + a_w^{\mathfrak n}, \quad
X= X^{(b)} + x^{\mathfrak t} + \frac {1}{g} x^{\mathfrak n},\quad
X^I = x^{I{\mathfrak t}} + \frac{1}{g} x^{I{\mathfrak n}},
\quad \theta^{\mathfrak n}= \theta^{\mathfrak t} +\frac{1}{\sqrt{g}} 
\theta^{\mathfrak n}\0
\ee
and likewise for the conjugate variables. For the ghosts we set
\be
c= g c^{\mathfrak t} + \sqrt g  c^{\mathfrak n}, \quad
\bar c= g \bar c^{\mathfrak t} + \frac {1}{\sqrt g} \bar c^{\mathfrak n}\,.
\0
\ee
These rescalings introduce a unit Jacobian in the path integral measure
of the non--zero modes, but they may produce a non-trivial factor due to
the presence of zero modes (see below).

After these rescalings the action becomes
\be
S=S_{sc}+Q_{\mathfrak n}+o\left(\frac{1}{\sqrt{g}}\right)\0\,,
\ee
where
\be
S_{sc}=
\frac{1}{\pi} \int_{{\cal C}_0} d^2w \,&\Tr& \left[
D^{(b)}_w x^{I{\mathfrak t}} D^{(b)}_{\bar w} x^{I{\mathfrak t}} 
+ D^{(b)}_w x^{{\mathfrak t}} D^{(b)}_{\bar w} \bar x^{{\mathfrak t}} 
 +i (\theta^{\mathfrak t}_s D^{(b)}_{\bar w} \theta^{\mathfrak t}_s 
+ \theta^{\mathfrak t}_c D^{(b)}_ w \theta^{\mathfrak t}_c)\right.\0\\
&&\left. +D^{(b)}_w a^{\mathfrak t}_{\bar w}
D^{(b)}_{\bar w} a^{\mathfrak t}_w + 
D^{(b)}_w \bar c^{{\mathfrak t}} D^{(b)}_{\bar w} c^{{\mathfrak t}}
\right]\label{scaction}
\ee
and $D^{(b)}$ represents the covariant derivative with respect to the 
background connection. $Q_{\mathfrak n}$ is the purely quadratic term in the 
$\phi^{\mathfrak n}$ fluctuations. The latter can be easily integrated over 
and, since they do not involve zero modes, give exactly 1, in accord with
supersymmetry (see Appendix~\ref{B}).

In conclusion, in the strong coupling limit we are left with the quadratic 
action (\ref{scaction}) over the Cartan modes.

\subsection{Lifting to the branched covering}\label{2.3}

Let us now show that the effective theory we obtained in the previous 
subsection corresponds to the Green--Schwarz superstring plus a free Maxwell
action on the worldsheet 
identified by the branched covering of the relevant background.
To this end we consider the quadratic action (\ref{scaction}) in the diagonal 
background: the covariant derivative $D^{(b)}_w$ becomes the simple derivative 
and the action becomes a free action
\be
S_{sc}=
\frac{1}{\pi} \int_{{\cal C}_0} d^2w \,&\Tr& \left[
\d_w x^{I{\mathfrak t}} \d_{\bar w} x^{I{\mathfrak t}} 
+ \d_w x^{{\mathfrak t}} \d_{\bar w} \bar x^{{\mathfrak t}} 
 +i (\theta^{\mathfrak t}_s \d_{\bar w} \theta^{\mathfrak t}_s 
+ \theta^{\mathfrak t}_c \d_ w \theta^{\mathfrak t}_c)\right.\0\\
&&\left. +\d_w a^{\mathfrak t}_{\bar w}
\d_{\bar w} a^{\mathfrak t}_w + 
\d_w \bar c^{{\mathfrak t}} \d_{\bar w} c^{{\mathfrak t}}
\right]\label{scaction1}.
\ee
Since all the matrices are diagonal we can rewrite this action in terms 
of the diagonal modes $\phi^{\mathfrak t} = \phi_{(1)},\ldots, \phi_{(N)}$:
\be
S_{sc}=
\frac{1}{\pi} \int_{{\cal C}_0} d^2w \,&\sum_{n=1}^N& \Big[
\d_w x^i_{(n)} \d_{\bar w} x^i_{(n)}
+i (\theta_{s(n)} \d_{\bar w} \theta_{s(n)} 
+ \theta_{c(n)} \d_ w \theta_{c(n)})\0\\
&& +\d_w a_{\bar w(n)}
\d_{\bar w} a_{w(n)} + 
\d_w \bar c_{(n)} \d_{\bar w} c_{(n)}
\Big]\label{scaction2}.
\ee
This is a theory of free fields on ${\cal C}_0$ and it is tempting to extend 
the action to ${\cal C}$ by just forgetting the punctures on the cylinder 
corresponding to the branch points. However this is not correct. The fields
$ x^i$ are not single--valued on the cylinder. For example, in the
${\mathbb Z}_N$ covering case, upon going around a branched point, each
$x^i$ is mapped to the adjacent one, and this is precisely the reason why 
they describe long strings, \cite{GHV}\cite{bbn}. Mathematically, this problem 
can be rephrased as follows: the fields 
in (\ref{scaction2}) are not section of bundles over ${\cal C}$ but they can 
be suitably combined to form sections of line bundles on the covering 
$\Sigma$ of ${\cal C}$. 

At this point it is worth spending a few words about {\it Hitchin systems}.
The Hitchin systems we are interested in are defined starting from a 
$U(N)$ vector bundle $V$ over ${\cal C}$, associated with the fundamental 
representation of $U(N)$. They consist of couples $(A,X)$ where $A$ is 
a gauge connection and $X$ a section of $End V \otimes K$, where $K$ is the
canonical bundle of ${\cal C}$, which satisfy (\ref{insteq1}) and
(\ref{insteq2}), \cite{hitchin1}. Such systems can be lifted to 
an $N$--branched covering of ${\cal C}$,\cite{hitchin2},\cite{mark},\cite{WD}.  
A remarkable feature of the lifting is the appearance on the branched covering
of a line bundle $L$ constructed out of $V$ and from which in turn $V$ can be 
reconstructed. In simple words the initial non--Abelian system can be
described by an equivalent Abelian system on the branched covering.
This is exactly the situation we are faced with when lifting our action to the 
branched covering. Without using rigorous mathematics, let us try 
to render this fact plausible by looking at the realization of 
local fields on a Riemann surface represented as a branched covering.
If $\Sigma$ is a branched covering of the cylinder ${\cal C}$, then there 
exists a projection map $\pi:\Sigma\,\,\to\,\,{\cal C}$
whose inverse image is N-valued. In our language this is simply
\be
\pi^{-1}:\, w\,\,\to\,\, \left(x_{(1)}(w),\dots,x_{(N)}(w)\right).
\ee
Suppose a local complex field $\tilde{\psi}$ is given on $\Sigma$;
applying the above construction $\tilde{\psi}$ can be represented
as an N-tuple $\left(\psi_{(1)}(w),\dots,\psi_{(N)}(w)\right)$
representing the field on each copy of the cylinder ${\cal C}$ 
that composes the covering $\Sigma$; the $\psi_{(i)}(w)$'s are related
by the appropriate monodromy properties along the curves ${\Re} w =const$. 

Going now back to the action (\ref{scaction2}), we will interpret any 
set of $N$ fields $(\phi_{(1)},\ldots, \phi_{(N)})$ in it as a 
unique field $\tilde \phi$ on the covering $\Sigma$. $\tilde \phi$ 
is locally a function
of a coordinate $z$ in $\Sigma$.  
From the point of view of $\Sigma$, the $w$ coordinate is locally
defined via an abelian differential $\omega=dw$ with imaginary periods which
is canonical, i.e. is fixed only by the complex structure
of the surface (see \cite{gidwol} and \cite{mand}).

Finally we can write the strong coupling action (\ref{scaction2}) as follows
\be 
&&S_{sc}^\Sigma= S^\Sigma_{GS} + S^\Sigma_{Maxwell},\label{GS+M}\\
&&S^\Sigma_{GS}=\frac{1}{\pi} \int_{\Sigma} d^2z \left( 
\d_z \tilde x^i\d_{\bar z} \tilde x^i
+i (\tilde\theta_{s} \d_{\bar z} \tilde\theta_{s} 
+\tilde \theta_{c} \d_z \tilde\theta_c)\right)\label{GS}\\
&&S^\Sigma_{Maxwell}= \frac{1}{\pi} \int_{\Sigma} d^2z\left(
g^{z\bar z}
\d_z \tilde a_{\bar z}
\d_{\bar z} \tilde a_{z} + 
\d_z \tilde {\bar c} \d_{\bar z}\tilde c\right)\,.\label{Max}
\ee
 In (\ref{GS})
a $\sqrt{\omega_z}$ (resp. $\sqrt{\omega_{\bar z}}$) factor 
has been absorbed in $\tilde{\theta_s}$ (resp. $\tilde{\theta_c}$)
which is a $(\frac{1}{2},0)$ (resp. $(0,\frac{1}{2})$) differential 
on $\Sigma$ and the metric in the Maxwell term is $g_{z \bar{z}}=
\omega_z\omega_{\bar z}$.

Summarizing, what we obtained in this subsection is that the strong coupling 
effective theory is given by the Green--Schwarz superstring action on the 
branched covering worldsheet plus a decoupled Maxwell theory on the same 
surface.

The fields in (\ref{GS}) are now sections of line bundles over  $\Sigma$,
i.e. well defined fields on the Riemann surface: for example $\tilde a$ is 
a section of the canonical bundle of $\Sigma$ and so on.

\subsection{String amplitudes}

Let us compute first, for simplicity, the vacuum to vacuum amplitude
of the SYM theory in the strong coupling limit in the background of a given
instanton. As we have already pointed out several times, this amplitude (up
to the vertex string insertions, see below) has a
string interpretation as the amplitude for the transition from the initial 
to the final string configuration described by the instanton. If this 
interpretation is correct this amplitude, to the leading order, should be
proportional to $g_s^{-\chi}$ where $\chi$ is the Euler characteristic
of the Riemann surface $\Sigma$, i.e. the covering surface introduced above.

What remains for us to do in order to evaluate this amplitude is to integrate
over the Cartan modes in the functional integral with action (\ref{GS})
(the non--Cartan modes have been integrated out above). Since the action is
free, the integration produces a ratio of determinants, which turns out to be 
a constant. However we have to take account of the
zero modes for the fields that have been rescaled (the unrescaled zero modes 
are irrelevant in this argument). The rescaled fields in ${\cal C}$
are the Maxwell and the ghost fields. The corresponding fields in $\Sigma$ will
be rescaled too
\be
\tilde a_z\,\to g\, \tilde a_z\, ,\quad \tilde a_{\bar z}\,\to g\, 
\tilde a_{\bar z}\, ,\quad
\tilde c\,\to g\,\tilde c\, ,\quad
\tilde {\bar c}\,\to g\, \tilde{\bar c}
\, .\label{rescalacc}
\ee

Therefore let us single out the Maxwell (plus ghost) partition function.
We will show that the decoupled $U(1)$ theory is there to generate the 
stringy factor $g_s^{-\chi}$ as a consequence of the rescaling
(\ref{rescalacc}), (on the Maxwell partition function, see also section~\ref{sec:3}).
 
In fact under this rescaling the Maxwell partition function ($a=a_z, \bar a=
a_{\bar z}$)
$$
Z^{\Sigma}_{Maxwell}=
\int{\cal D}\left[\tilde{a},\tilde{\bar a},\tilde{c},\tilde{\bar c}\right]
\,\, e^{-S^{\Sigma}_{Maxwell}(\tilde{a},\tilde{\bar a},\tilde{c}\,,
\tilde{\bar c})}
$$ 
rescales with a factor depending on the zero modes. Roughly speaking, 
what happens is that the above integral is interpreted as the ratio 
$({\det}'\square_c)/({{\det}' \square_a})$, where $\square= \d \bar \d$ 
denotes the quadratic operator in the action, and $'$ means that the zero 
modes have been excluded from the computation of the regularized determinants; 
since we have rescaled 
the measure there will arise a factor of $g$ at a power equal to the unbalance 
of the zero modes \footnote{A more precise account of this point can be found 
in 
\cite{witten1} where an analogous calculation of the Maxwell partition function 
was carried out.}. The problem is therefore to count the latter. 
As for the ghost fields which are scalars, the only 
zero modes of the $\bar \d$ operator on $\Sigma$ is the constant. The
zero modes of the Maxwell fields correspond to the holomorphic 
differential on $\Sigma$. If $\Sigma$ were a closed Riemann surface of genus
$h$, their number would be $h$. Their counting in the present case is not 
completely standard as $\Sigma$ is actually a Riemann surface with boundaries 
(representing the in-- and out-- strings).
A way to do the counting is to construct the double $\hat\Sigma$ of $\Sigma$:
$\hat \Sigma$ has genus $\hat h=2h +b-1$, where $b$ is the number of 
boundaries, and admits an anticonformal involution with the set of fixed points
corresponding exactly to the boundary of $\Sigma$. We can count now the number 
of analytic differential on $\Sigma$ that extend to $\hat \Sigma$, that is
the so--called analytic Schottky differentials \cite{ahl}: their number is
$\hat h$. Therefore the overall unbalance of zero modes (including the ghosts)
is $\hat h -1= 2h + b -2$, which is exactly the opposite of the Euler number
of $\hat\Sigma$. An equivalent way of deriving this result is to use the
Gauss--Bonnet theorem on $\hat\Sigma$ and noting that, due to the involution,
the integral of the curvature over $\Sigma$ is one half of the total 
contribution. 

Finally the factor in front of the vacuum to vacuum amplitude
will be $g^{-2h-b +2} = g_s^{2h+b-2}$. The exponent of $g$ is precisely the 
Euler characteristic of $\Sigma$, as we wanted to prove.

In order to appreciate exactly what we have just computed we must now specify
what it corresponds to in string interaction theory. In this 
sense the amplitude we have just computed in the strong coupling limit is a 
basic amplitude but, of course, an incomplete one. 

First of all, real string amplitudes should contain 
vertex string insertions, i.e. should be correlators of the vertex
operators corresponding to the various in-- and out-- (super)strings.
In this regard we simply remark that such vertex operators are constructed
in terms of the string fields $\tilde x^i$ and $\tilde\theta$ (and, possibly,
the string ghosts and superghosts), therefore
the treatment of the non--Cartan modes above is not affected and the discussion
of the zero modes of the Maxwell sector is unchanged. Therefore the scaling 
factor $g_s^{-\chi}$ is left unchanged too.

Moreover, in order to obtain complete amplitudes, we must still integrate 
over the moduli of the
Hitchin systems, i.e. over the inequivalent BPS instantons that interpolate 
between a given initial and a given final state. If we want to
implement such more advanced stage of calculation, we have to take into 
account in the measure some Jacobians that are produced by the various field
splittings we have considered above.
Following \cite{gersa}, the background/fluctuations splitting of the fields
in the path integral generates a Jacobian $J_{b/f}$. In an analogous way
also the Cartan/non--Cartan splitting gives rise to another Jacobian factor 
$J_{C/nC}$. These factors are easily seen to depend only on the Cartan modes
of $X$ and $\theta$.
This, in particular, implies the validity of the procedure we used for the
integration over the non--Cartan modes.

The total partition function in the strong coupling limit is therefore
\be
Z_{sc}=\int_{{\cal M}_{sc}}dm~ g_s^{-\chi}
\int {\cal D}[\tilde{x},\tilde{\theta},\tilde{a},\tilde{c}] J_{b/f}J_{C/nC} 
\e^{-S_{GS}-S_{Maxwell}}\,,
\ee
where ${\cal M}_{sc}$ is the strong coupling instanton moduli space. But 
we know what this is: the 
strong coupling limit of each BPS configuration is completely 
identified by the relevant branched covering. This means that the integral
over the background instantonic configurations corresponds in the strong 
coupling to the sum over all inequivalent Riemann surfaces with appropriate 
boundaries. What is important to stress here is the 
$g_s^{-\chi}$ factor in front of the path integral at fixed genus.
 
At this point the situation is clear: Matrix String Theory in the strong 
coupling limit looks very much like IIA theory. We would like to say `is'
instead  of `looks like'; however a complete proof of the 
equivalence of the two theories requires showing that  
the above Jacobian factors give rise to the full superstring measure
in the path integrals over the surfaces \cite{dhofo}; this in turn requires
a manageable representation of the moduli space of the Hitchin systems
in the strong coupling limit.

\section{Comments}\label{sec:3}

In this final section we would like to make a few comments, in part to complete
the previous presentation and in part to outline future developments.

The first comment concerns the explicit breaking of the $SO(8)$ 
invariance of (\ref{eSYM})
in the background: we recall that, at the moment of choosing a fixed BPS 
background, we set $X^i\neq0$ only for $i=1,2$. As one can see, in
the strong coupling limit no remnant of this explicit breaking survives.
However the problem of $SO(8)$ symmetry breaking may arise in non--zero orders
of the $1/g$ expansion. In such a case $SO(8)$ invariance is recovered
by summing over the instantons lying along all couples of the eight directions.

Next let us briefly discuss the relation of the results of this paper
with the large $N$ limit, \cite{BFSS}. It must be stressed that 
what we have found above is in agreement with the present 
understanding of the M(atrix) theory \cite{suss}.
In fact in order to see string worldsheets, and make computations with them, 
we do not need to take the 
large $N$ limit: they are clearly visible and effective for finite $N$ and 
the large $N$ limit does not add anything to such visibility.
Of course the $U(N)$ theory describes only a limited amount of string processes
due to the fact that the total content of each asymptotic sector, given by
a definite long string configuration  
$(1)^{n_1}\cdot(2)^{n_2}\cdot\dots\cdot(s)^{n_s}$,
has finite size $N=\sum_{i=1}^s in_i$. The full infinite set of possible
string processes can be truly obtained only in the large $N$ limit.

We would like to end the paper with some comments concerning the free
Maxwell action and the related partition function, which appears in the strong 
coupling limit. Above Maxwell field $\tilde a$
on $\Sigma$ appears as a `small' fluctuation. We would like to 
be more precise about this point. 

Following \cite{witten2}, the quantization of the Maxwell sector is 
straightforward, 
\be
Z^{\Sigma}_{Maxwell}({\tt m})
=
\sum_{R}\left({\rm dim}\,\, R\right)^{\chi_\Sigma}
\prod_{s=1}^n \chi(R)_{{\tt m}_i}
e^{-\frac{A_\Sigma}{2}C_2(R)}
\, ,\label{wittenpf}
\ee
where ${\tt m}=({\tt m}_1,\dots,{\tt m}_n)$ are the periods of the $U(1)$ 
connection around the boundaries of $\Sigma$, $R$ is any irreducible $U(1)$ 
representation, $\chi(R)_{{\tt m}_i}$ is the relative character,
$A_\Sigma=\int_\Sigma \sqrt{g} d^2z$ is the area of $\Sigma$ in the 
branched covering metric and $C_2(R)$ is the Casimir of $R$.

Explicitly evaluating (\ref{wittenpf}), we get
\be
Z^{\Sigma}_{Maxwell}({\tt m})
&=&
\sum_{l\in{\Z}}
e^{il|{\tt m}|}
e^{-\frac{NA_{\cal C}}{2}l^2}\0\,,\\
\ee
where $|{\tt m}|=\oint_{\partial\Sigma}\tilde{a}=\int_{\Sigma} \tilde f$ 
and $\tilde f=d\tilde a$ is the curvature of $\tilde a=
\tilde a_zdz+\tilde a_{\bar z}d{\bar z}$.
We assume $|{\tt m}|$ to vanish, and this is our definition of 
Maxwell fluctuation in this paper. This means in particular that
the Maxwell field is just a section of the canonical line bundle,
as we assumed implicitly above while doing zero--modes counting.
With this assumption the Maxwell factor is the constant
$$
Z^{\Sigma}_{Maxwell}=\sum_{l\in{\Z}}
e^{-\frac{NA_{Cyl}}{2}l^2}\, ,$$
which goes to $1$ in the large $N$ limit.

We would like however to point out that the Maxwell theory on $\Sigma$ admits
other sectors in which $|{\tt m}|\neq 0$. These non--trivial sectors
have not been considered in 
the present paper. However their inclusion in the theory looks extremely
interesting. We recall that, in a non--interacting regime (trivial 
background) of MST, the number of units of quantized electric flux is 
identified with the D0--brane charge, \cite{GRT,BSS}. 
The number $|{\tt m}|$ may have to do with the conserved D0--brane charge
involved in the interaction.
Related to this is the remark that, although the Maxwell sector is 
completely decoupled from the 
string sector of the theory in the strong coupling limit, ${\cal O}(1/g)$
and higher corrections to the effective action will bring into the game 
interaction terms representing non--perturbative contributions to string 
theory.

\appendix

\section{Solutions for the Toda dressing factors}\label{A}

In this Appendix we discuss the properties of the
Toda dressing factor $e^{u\cdot H}$ in the classical instanton solutions
in the strong coupling limit. We recall that this factor appears in the
dressing matrix $Y$. With respect to \cite{bbn} we introduce here a 
convenient modification: instead of matrices $Y,J, U$ in
$GL(N,{\mathbb C})$ we will 
consider matrices $Y,J,U$ in $SL(N,{\mathbb C})$; simultaneously the matrices $H$ and 
${\cal H}$  will be traceless. Any result can be obtained directly from 
\cite{bbn} by simply dividing by the corresponding 
determinants or subtracting the corresponding traces. The reason for this is 
that the determinant/trace factors introduce a useless complication for our 
analysis. With this understanding we will use the same symbols as in \cite{bbn}:
in particular
$(H_i)_{kl} = (\delta_{i,k}-\delta_{i,k+1})\delta_{kl}$, $i=1,...N-1$ and 
$k,l=1,...,N$, and ${\cal H}= Diag((1-N)/2N,(3-N)/2N,...,(N-1)/2N)$.

In the case of the ${\mathbb Z}_N$ covering $X^N -a=0$ the `fields' $u\cdot H$ 
satisfy the Toda--like equations
\be
\d_\zeta \d_{\bar \zeta} (u\cdot H)+ g^2 \left[e^{-u\cdot H}{\cal P} 
e^{u\cdot H}, {\cal P}^\dagger\right]=
\pi {\cal H}\delta(a)(\d_\zeta \bar a)(\d_{\bar \zeta} a)\,,.\label{todaeq}
\ee
where $\frac {\d \zeta }{\d w} = \bar a ^{1/N}$. In particular, in the case 
$N=2$, we have only one field $u$ satisfying
\be 
\d_\zeta \d_{\bar \zeta} u - 2g^2 \sinh~ u= - \frac{\pi}{4} \delta (a) 
(\d_\zeta \bar a)(\d_{\bar \zeta} a)\label{sinh}\,.
\ee 
This means that $u$ must satisfy the sinh-Gordon equation with the boundary 
condition
\be
u \sim - \frac{1}{4} \ln |a|, \quad\quad a\sim 0\label{boundary}\,.
\ee
Let us suppose, for simplicity, that $\bar a = z-z_0 $, where $z_0$ is a 
point away from the origin of the $z$--plane, and $z= e^w$. This
corresponds to the case of $N$ 
closed string joining to form one long string, see \cite{bbn}. Now we want to 
single out, for this simple case, solutions of  (\ref{sinh}) with the right 
asymptotic behaviour to interpolate between the in- and out- string states. 
First we rescale $\zeta\to \sqrt 2 g \zeta$, so that the sinh--Gordon
equation takes the standard form
\be
\d_\zeta \d_{\bar \zeta} u - \sinh\, u=0 \label{sinh1}\,.
\ee
We do not know an exact solution of this equation satisfying the boundary
condition (\ref{boundary}) and vanishing at the origin and at infinity
of the $z$--plane. We can therefore proceed in two ways: find a numerical 
solution or an approximate analytical one. Let us do the latter first.

Recalling that
\be
\frac {\d \zeta}{\d z}= \sqrt 2 g \frac {\sqrt {z-z_0}}{z}\0\,,
\ee
the approximate expression of $\zeta$ in terms of $z$ is
\be
&& \zeta \sim  \frac  {2\sqrt 2g}{3 z_0} ( z-z_0)^{3/2}, \quad {\rm for}\quad 
z\sim z_0\0\\
&& \zeta \sim \frac {\sqrt 2 g}{2} \sqrt {z} , \quad {\rm for}\quad 
|z|>>|z_0|\0\\
&& \zeta \sim \sqrt 2 g \sqrt{-z_0}\,\ln\,z,\quad {\rm for}\quad|z|<<|z_0|\0\,.
\ee
If these were the exact expressions for $\zeta$, we could
consider spherically symmetric solutions of (\ref{sinh1}), i.e. solutions 
depending only on $r= |\zeta|$. For them eq.(\ref{sinh1}) takes the form
\be
\d_r^2 u+ \frac {1}{r}\d_r u = 4 \sinh\, u\label{painleve}
\ee
This is a form of the Painlev\'e III equation. The general form of the 
solutions of this equation are known, see \cite{IN} and references therein.  
Let us select the class of solutions with the following asymptotic behaviour:
\be
&& u(r) \sim \alpha \,\ln\,r + \beta, \quad\quad r\to 0 ,\quad |\alpha|<2 
 \label{asym1}\\
&& u(r) \sim \gamma r^{-1/2} e^{-2 r},\quad\quad r\to \infty\label{asym2}\,.
\ee
The constants $\beta$ and $\gamma$ must be fine--tuned to $\alpha$ in order 
to give rise to smooth solutions. However, in our case, we are not interested 
in the actual value of 
$\beta$ and $\gamma$, therefore we can always adjust the parameters in such a 
way as
to have a smooth solution. The only serious problem might come from the bound 
$|\alpha|<2$; however the asymptotic behaviour (\ref{asym1}) is in our case
\be 
u \sim  -\frac{1}{6} \ln\, r\0
\ee
therefore the bound is satisfied. 

Let us study now the properties of the solution. In the various regions 
we have the following asymptotic expressions:
\be
&& r \sim  \frac  {2\sqrt 2g}{3 |z_0|}| z-z_0|^{3/2}, \quad {\rm for}\quad 
z\sim z_0\0\\
&& r \sim \frac {\sqrt 2 g}{2} \sqrt {|z|} , \quad {\rm for}\quad 
|z|>>|z_0|\label{asymp}\\
&& r \sim \sqrt 2 g \sqrt{|z_0|}\ln |z|,\quad {\rm for}\quad,|z|<<|z_0|\,.
\ee
From this we see that, when, at fixed finite $g$, $z$ is near the origin and 
far to infinity, the solution tends to zero. The convergence to zero is more 
rapid the larger $g$ is,
the slope in $g$ being of negative exponential type. Looking now at the first 
equation (\ref{asymp}), we see that, even if we sit near $z_0$, we may still 
fall in the regime ($r$ large) in which the solution is extremely small, 
provided $g$ is large enough. In other
words, for large $g$ the solution shrinks around $z_0$, and, in the 
$g\to \infty$ limit
it becomes spike--like with support at $z=z_0$. We can say that, if we exclude
a neighborhood of $z_0$ of size proportional to $(g)^{-2/3}$, the solution 
decreases to zero more rapidly than any power of $1/g$.

We recall that the spherically symmetric solution is not the exact solution, but
only an approximate one. However we expect the general behaviour of the 
true solution to be essentially similar, i.e. that it shrinks very rapidly 
around $z_0$ as $g\to \infty$. This is confirmed by the numerical solutions
shown in fig.~\ref{f1}.

\FIGURE[t]{%
$(a)$\epsfig{file=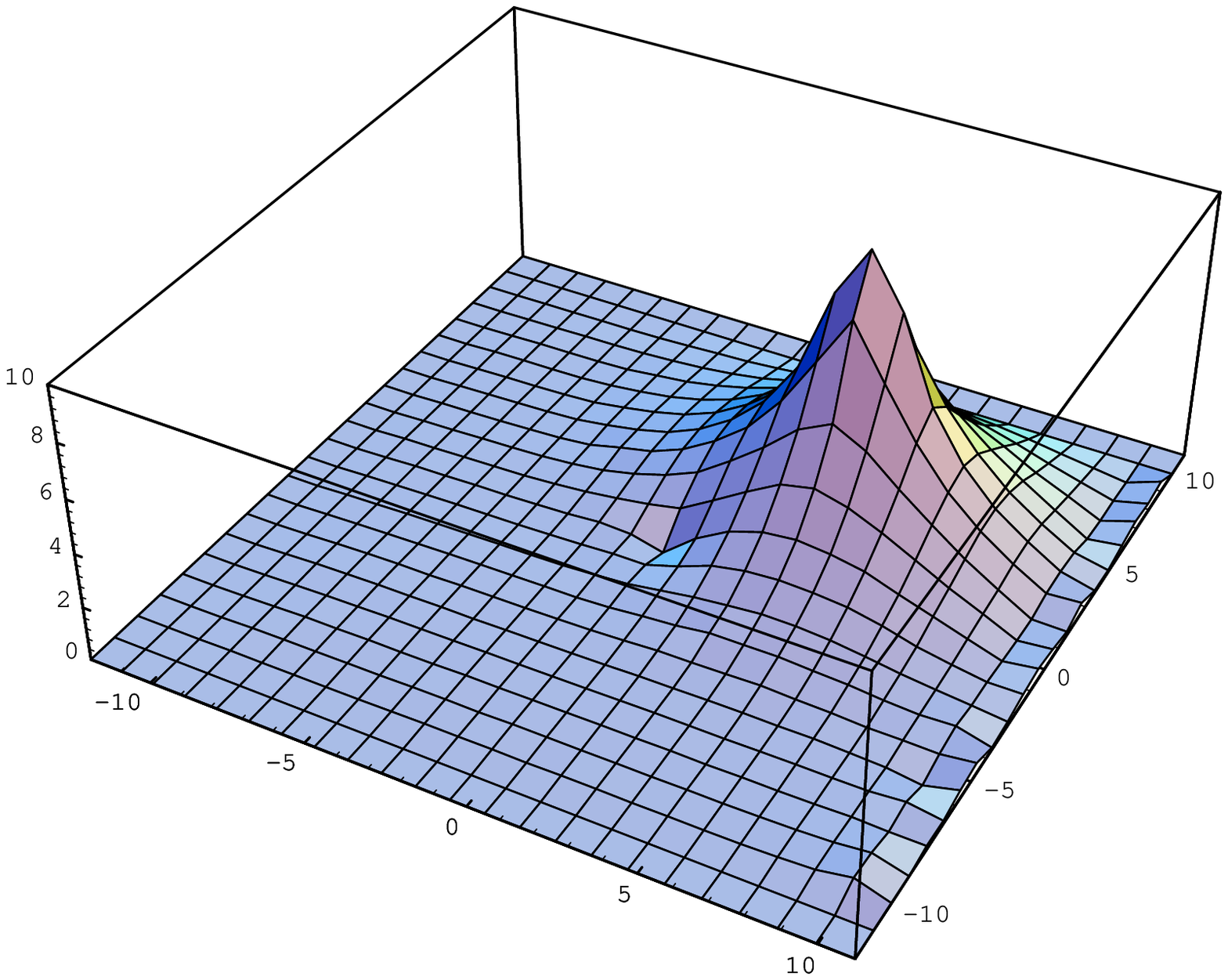,width=15.5em}\hspace{2em}
$(b)$\epsfig{file=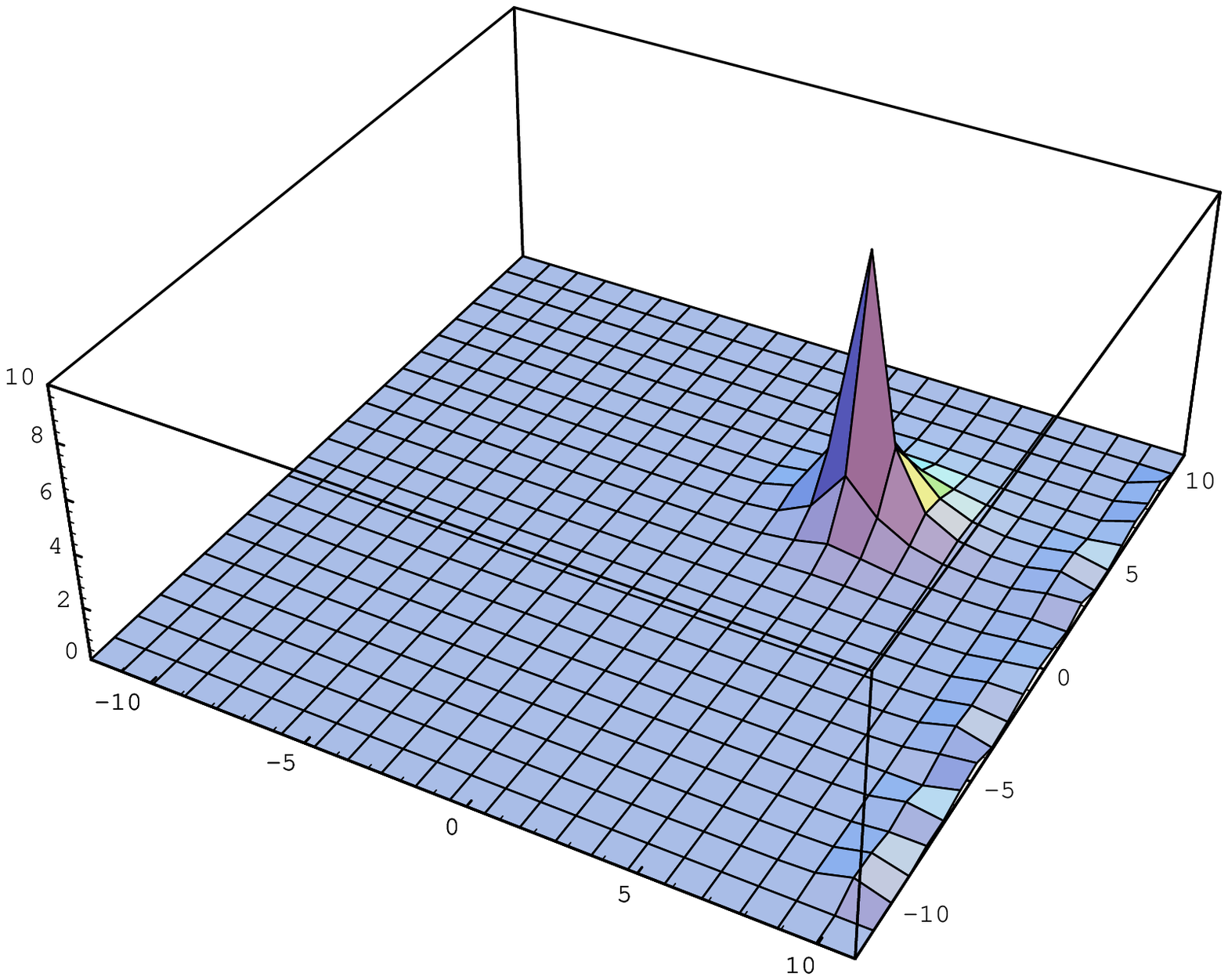,width=15.5em}%
\caption{Numerical solutions of eq.~(\ref{sinh}) for $(a)$ small and
$(b)$ large $g$, respectively. For the Bullough-Dodd case, eq.~(\ref{BD}), the
solutions are almost identical.\label{f1}}}

We can easily extend the previous analysis to the case in which $a$ contains
several distinct zeroes. Simply find suitable approximate expressions for 
$\zeta$
near the zeroes of $a$ and apply the previous approximate analysis. The 
conclusion will be that the solution shrinks very quickly around the branch
points as $g\to \infty$.

Let us briefly consider the case $N=3$. We have two fields $u_1,u_2$ obeying
the equations
\be
&&\d_\zeta \d_{\bar \zeta} u_1 + g^2 \left(e^{-u_1-u_2}- e^{2u_1-u_2}\right)
= -\frac{\pi}{3} \delta(a)~ \d_{\bar \zeta}a~\d_{ \zeta}\bar a\0\\
&&\d_\zeta \d_{\bar \zeta} u_2 + g^2 \left(e^{-u_1-u_2}- e^{2u_2-u_1}\right)
= -\frac{\pi}{3} \delta(a)~ \d_{\bar \zeta}a~\d_{ \zeta}\bar a\0\,.
\ee
Subtracting one from the other we get an equation which is identically 
satisfied with $u_1=u_2$. The remaining equation for $u=u_1=u_2$ is
\be
\d_\zeta \d_{\bar \zeta} u + g^2 \left(e^{-2u}- e^{u}\right)
= -\frac{\pi}{3} \delta(a)~ \d_{\bar \zeta}a~\d_{ \zeta}\bar a\label{BD}\,.
\ee
This is known as the Bullough--Dodd equation. The features of the
solutions, with the present boundary conditions, are essentially 
the same as for the sinh--Gordon equation: for $g \to \infty$ they 
shrink around the zeroes of $a$. The numerical solutions do not differ 
significantly from the sinh--Gordon case. 

We will not discuss the $N>3$ case here.

\section{Non--Cartan fluctuations}\label{B}

In  this Appendix we briefly discuss the $Q_{\mathfrak n}$ term of the
strong coupling action, i.e. the quadratic terms in the non--Cartan modes.
$Q_{\mathfrak n}$ has the form
\be
Q_{\mathfrak n}= \frac {1}{\pi} \int d^2w \Tr \left[ \bar 
x^{\mathfrak n}{\cal Q} x^{\mathfrak n}+  
x^{I{\mathfrak n}}{\cal Q} x^{I{\mathfrak n}}+ 
a_{\bar w}^{\mathfrak n}{\cal Q} a_w^{\mathfrak n}+
\bar c^{\mathfrak n}{\cal Q} c^{\mathfrak n}+
i (\theta_s^{\mathfrak n}, \theta_c^{\mathfrak n}) {\cal A}
\left(\matrix{\theta_s^{\mathfrak n}\cr \theta_c^{\mathfrak n}\cr}\right) 
\right]\label{Qn}\,,
\ee
where 
\be
{\cal Q} = {\rm ad}_{{\bar X}^\circ}\cdot {\rm ad}_{ X^\circ}+
 {\rm ad}_{a_{\bar w}^t}\cdot {\rm ad}_{a_w^t} +
 {\rm ad}_{x^{It}}\cdot {\rm ad}_{x^{It}}\0
 \ee
and
\be
{\cal A} =\left( \matrix { i {\rm ad}_{a_{\bar w}^t} & \gamma_i 
{\rm ad}_{X^{\circ i}} \cr
\tilde \gamma_i {\rm ad}_{ {\bar X}^{\circ i}} & 
i {\rm ad}_{a_{ w}^t}\cr}\right)\0\,.
\ee
In the path integral we can now integrate over the non--Cartan modes and obtain
a ratio of determinants of ${\cal A}$ and ${\cal Q}$. Since these operators
do not have zero modes the calculation is elementary. The integration over
$a^{\mathfrak n}$ and the conjugates exactly cancels the integration over 
$c^{\mathfrak n}$ and the conjugates. What remains is a ratio 
$\left((\det {\cal A})^{16}/(\det {\cal Q})^8\right)^{N^2-N}$. 
The expression of the numerator is 
formal: one should understand ${\det {\cal A}}$ as $\sqrt {\det (-{\cal A}
{\cal A}^\dagger)}$. But ${\cal A}{\cal A}^\dagger= {\cal A}^\dagger 
{\cal A}= - {\cal Q}$. Therefore the net result of integrating over the 
non--Cartan modes is 1. This is the result expected from supersymmetry in the 
absence of zero modes.

\acknowledgments
We would like to acknowledge valuable discussions we had 
with E.\ Aldrovandi, D.\ Amati, C.S.\ Chu, G.\ Falqui, F.\ Morales,
C.\ Reina and A.\ Zampa. This research was partially supported by EC
TMR Programme, grant FMRX-CT96-0012, and by the Italian MURST for the
program ``Fisica Teorica delle Interazioni Fondamentali".

\end{document}